# Discovery of a very extended X-ray halo around a quiescent spiral galaxy – the "missing link" of galaxy formation

Kristian Pedersen[a], Jesper Rasmussen[b], Jesper Sommer-Larsen[a], Sune Toft[c], Andrew J. Benson[d], Richard G. Bower[e]

[a]Dark Cosmology Centre, Niels Bohr Institute, University of Copenhagen, Juliane Maries Vej 30, DK-2100 Copenhagen, Denmark; kp@astro.ku.dk, jslarsen@astro.ku.dk

[b]School of Physics and Astronomy, University of Birmingham, Edgbaston, Birmingham, B15 2TT, UK; jesper@star.sr.bham.ac.uk

[c]Department of Astronomy, Yale University, P.O. Box 20810, New Haven, CT 06520-8101, USA; toft@astro.yale.edu

[d]Astrophysics, University of Oxford, Keble Road, Oxford, OX1 3RH, England, UK; abenson@astro.ox.ac.uk

[e]Institute for Computational Cosmology, Physics Department, University of Durham, South Road, Durham DH1 3LE, England, UK; r.g.bower@durham.ac.uk

Corresponding author: Kristian Pedersen. Phone: +45 3532 5932. Fax: +45 3532 5989

## Abstract

**Hot gaseous haloes surrounding galaxies and extending well beyond the distribution of stars are a ubiquitous prediction of galaxy formation scenarios. The haloes are believed to consist of gravitationally trapped gas with a temperature of millions of Kelvin. The existence of such hot haloes around massive elliptical galaxies has been established through their X-ray emission. While gas out-flowing from starburst spiral galaxies has been detected, searches for hot haloes around normal, quiescent spiral galaxies have so far failed, casting doubts on the fundamental physics in galaxy formation models. Here we present the first detection of a hot, large-scale gaseous halo surrounding a normal, quiescent spiral galaxy, NGC 5746, alleviating a long-standing problem for galaxy formation models. In contrast to starburst galaxies, where the X-ray halo can be powered by the supernova energy, there is no such power source in NGC 5746. The only compelling explanation is that we are here witnessing a galaxy forming from gradually in-flowing hot and dilute halo gas.**

## 1. Introduction

There are several indications that the formation of spiral galaxies, such as the Milky Way, is persisting to the present day, e.g. in-fall of high-velocity clouds towards the Milky Way disc (Wakker et al. 1999), and the need for a continuous supply of low-metallicity gas to explain the metallicity distribution of stars in the Solar neighbourhood (Pagel 1997). This is incorporated into current galaxy formation models which are calibrated to reproduce the optical and infrared emission from present day galaxies (Kaufmann et al. 1999, Cole et al., 2000, Bell et al. 2003). It has long been realised that a key test of these models for spiral galaxies is the luminosity and spatial distribution of X-ray emission from putative hot halo gas (Spitzer 1956) which is predicted to be cooling via thermal emission and gradually flowing into the galaxy potential even at the present epoch (White & Frenk 1991, Benson et al. 2000). The X-ray luminosity of the hot halo is predicted to increase strongly with the mass of the galaxy and the halo of the most massive spiral galaxies should be detectable with current X-ray instruments (Toft et al. 2002). However, due to the non-detection of hot haloes around spiral galaxies (Benson et al. 2000) the generic galaxy formation scenario has been questioned (Binney 2004), and it has been suggested that spiral galaxies of total mass less than a few times $10^{11}$ Solar masses ($M_\odot$) form primarily through in-fall of cold gas, hence showing no detectable X-ray halo (Binney 2004, Birnboim & Dekel 2003). Also, Lyman-α emission, from cold gas falling into massive dark matter haloes ($M \approx 10^{12}$-$10^{13}$ $M_\odot$) at intermediate redshift ($z$≈3) has recently been reported (Weidinger et al. 2004, Bower et al. 2004). On the other hand, the recent detection of a warm-hot phase of the intergalactic medium (Nicastro et al. 2005) shows the presence of a reservoir of hot and dilute gas at galactic distances $\sim 1$ Mpc . Furthermore, absorption of the OVI line in quasar spectra (Wakker et al. 2004), and the "head-tail" structure of $\sim 20\%$ of the high-velocity clouds in the halo of the Milky Way (Brüns et al. 2000) provide circumstantial evidence that the Milky Way is surrounded by an extended hot halo.

## 2. X-ray observations

In order to make a stringent test of current galaxy formation models we conducted a targeted study of the most promising candidate spiral galaxy for detecting halo X-ray emission, NGC 5746. We also studied a similar, but less massive galaxy, NGC 5170, as a test of our procedure. These galaxies were selected so as to maximise the expected halo soft X-ray flux and to minimize contaminating X-ray emission from other sources. The galaxies are massive and nearby (NGC 5746 is an SBb galaxy at a distance of 29.4 Mpc and has a circular velocity of 307±5 km/s (Tully 1988), where the circular rotation velocity is determined at 2.2 times the disc scale length, and NGC 5170 is an Sc galaxy at a distance of 24.0 Mpc (Tully 1988) and has a circular velocity of 250±5 km/s, Kregel et al. 2004). Both galaxies are quiescent, showing no signs of either starburst activity (star formation rates of 1.2 $M_\odot$ $yr^{-1}$ and 0.5 $M_\odot$ $yr^{-1}$ for NGC 5746 and NGC 5170 respectively, derived from their IRAS 12-100 μm luminosity, Moshir et al. 1990), interaction with other galaxies, or an active galactic nucleus. The discs of the galaxies are viewed almost perfectly edge-on, and the galaxies are situated more than 30° from the Galactic plane.

NGC 5746 and NGC 5170 were observed by the *Chandra X-ray Observatory* with the ACIS-I array in Very Faint mode on April 11-12, 2003 and May 18, 2003, respectively. The data were reprocessed using the `acis_process_events` task in CIAO version 3.02 and standard screened for bad detector pixels, non-X-ray events in the detector, and periods of high background, resulting in 36.8 ks (NGC 5746) and 33.0 ks (NGC 5170) of effective exposure time. Point sources were detected by the `wavdetect` procedure in CIAO and masked out in the further analysis using their `wavdetect` 3σ detection ellipses.

The background level was obtained in a two-step procedure (Rasmussen & Ponman 2004). First, the local background was determined from an annulus 5'-7' from the center of NGC 5746 (immediately outside the circle in Fig. 1). Next, the background from this detector region obtained from many combined blank sky background exposures (Markevitch 2005) was measured. Finally, from the combined blank sky exposures the background in the same detector region as used for extracting the halo emission (bounded by the circle and the ellipse in Fig. 1) was extracted, adding the difference between the local background and the blank sky background as measured in the annulus. The background in the spectral analysis was also extracted in this way and spectral responses weighted with the spectrum of the halo emission were used.

## 3. The very extended X-ray halo around NGC 5746

Diffuse, soft X-ray emission extending more than 20 kpc from the stellar disc was detected around NGC 5746, see Fig. 1. A total of about 200 net counts in the 0.3-2 keV band were detected from the halo of NGC 5746, corresponding to a 4.0 sigma detection. The same observing technique and data analysis revealed no diffuse emission around the less massive galaxy NGC 5170.

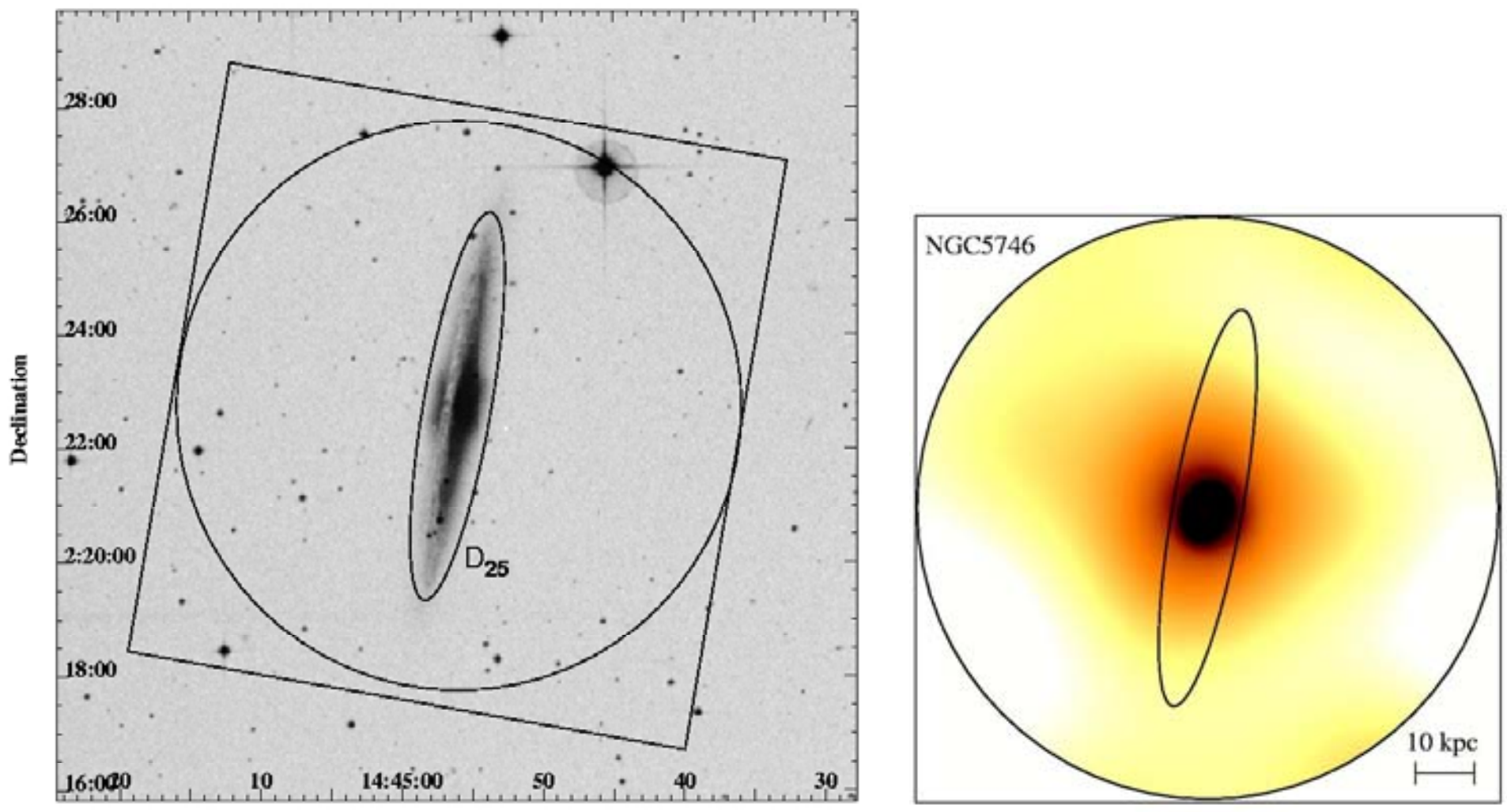

**Figure 1** Optical image of the Chandra ACIS-I field of view and X-ray images of NGC 5746. Left: Digitized Sky Survey image of NGC 5746. Right: Adaptively smoothed, exposure corrected, and background subtracted Chandra image where detected point sources have been removed. Prior to smoothing, each point source region was refilled by a Poissonian pixel value distribution sampled from the distribution of pixel values in the surrounding region. The outermost diffuse emission is significant at the 3 sigma confidence level, increasing to the 5 sigma confidence level near the disc. The circle outlines the outer boundary of the area within which the halo X-ray spectrum was extracted. The inner boundary was chosen as the ellipse outlining the *B*-band 25 mag arcsec$^{-2}$ isophote, $D_{25}$, shown in the images. The square marks the region for which the X-ray surface brightness distribution perpendicular to the disc was measured (see Fig. 2).

The morphology of the X-ray halo is difficult to map given the magnitude of the detected signal. We therefore made a one-dimensional parameterisation of the spatial distribution of the X-ray emission by extracting the number of counts in slabs parallel to the disc (see Fig. 2). The diffuse X-ray emission is fairly symmetrical around the disc and it is detected out to at least 20 kpc from the disc on both sides. The spatial extent of the X-ray halo of NGC 5746 considerably surpasses the extent of diffuse off-disc X-ray emission seen in any other quiescent spiral galaxy (Strickland et al. 2004a). The profile is well fitted with a power law plus a constant background $S(z) = S_0 \,|\, z \,|^{-\Gamma} + B_0$ with best fit parameters $\Gamma = 0.68^{+0.42}_{-0.35}$, $S_0 = 0.42^{+0.14}_{-0.07} \times 10^{-6}$ counts arcsec$^{-2}$, $B_0 = 0.31^{+0.03}_{-0.09} \times 10^{-6}$ counts arcsec$^{-2}$, where $z$ is the distance perpendicular to the disc in kpc, measured from the disc mid-plane. Quoted errors here and in the following are $1\sigma$ confidence. Since (as expected, see below), no statistically significant diffuse emission is detected around the less massive spiral galaxy NGC 5170, using exactly the same approach, we can rule out the possibility that the diffuse X-ray emission around NGC 5746 is due to instrumental artefacts or systematic errors in our data analysis.

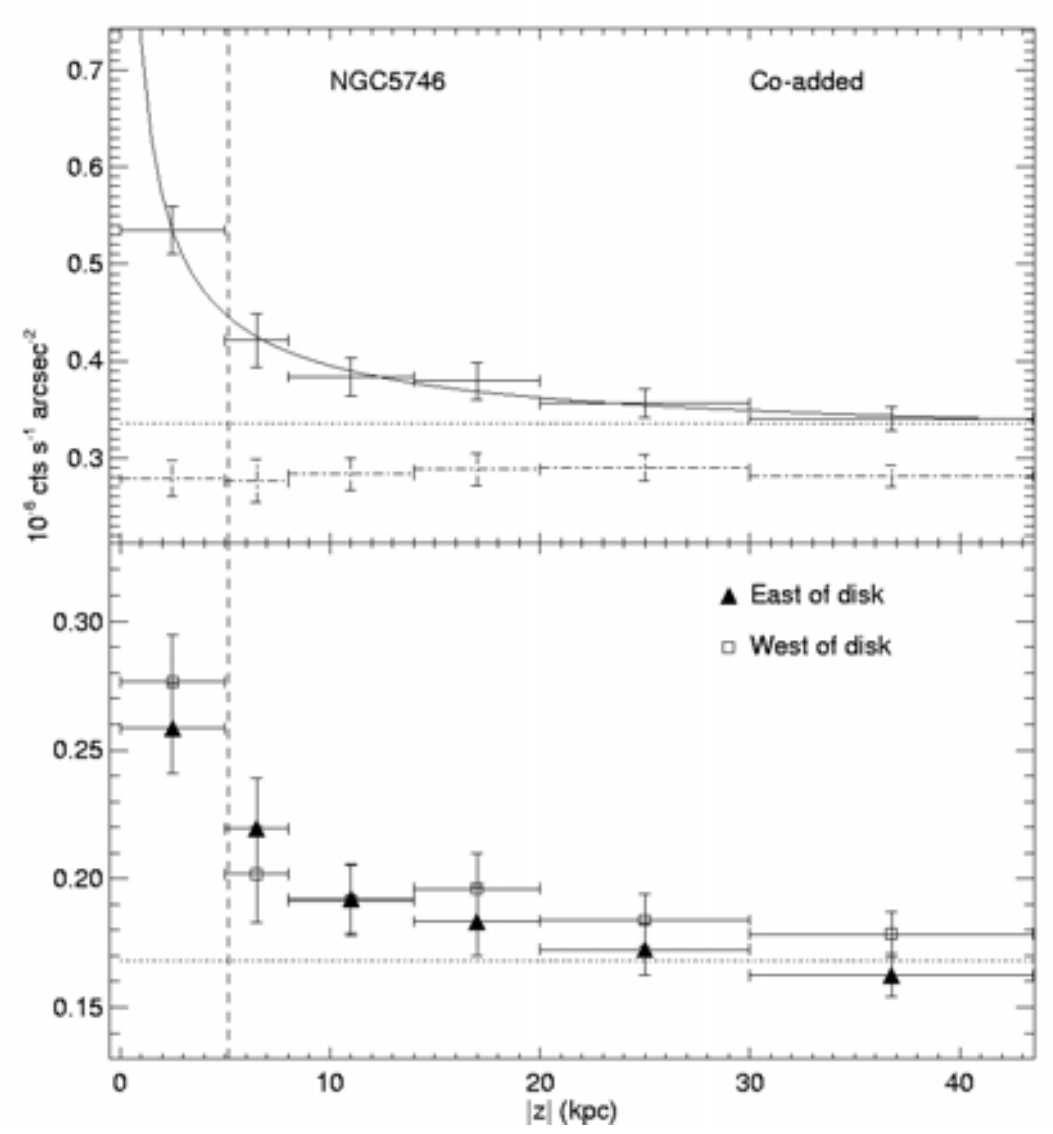

**Figure 2** X-ray surface brightness of diffuse 0.3-1.5 keV emission from NGC 5746. The X-ray surface brightness was measured in 90 kpc wide slabs parallel to the disc. The distance to the disc mid-plane, $z$, is measured perpendicular to the disc. Point sources and chip gaps were masked out. The vertical dashed lines indicate the extent of $D_{25}$ (see Fig. 1). The horizontal dotted lines mark our best estimate of the background level (see text). The dot-dashed points show the measured background level from blank-sky data being about 20% lower than the background in this direction on the sky. Top: Profile for emission co-added from both sides of the disc. The solid line shows the best fit power law plus a constant background profile. Bottom: Profiles for the Eastern side and Western side of the disc, respectively.

The 0.3-5 keV spectrum of the diffuse X-ray emission, excluding emission from the disc (see Fig. 1), is well fitted by a thermal plasma model with temperature $T = 6.5^{+2.1}_{-2.3} \times 10^6$ K, and a 0.3-2 keV luminosity $L_X = 4.4^{+3.0}_{-1.5} \times 10^{39}$ erg s$^{-1}$ (corrected for Galactic absorption) for a plasma with a metallicity 0.2 times the Solar value ($\chi^2 = 9.77$ for 9 degrees of freedom). The metallicity is not well constrained, but the temperature and the luminosity remain within the quoted statistical uncertainties for all sub-Solar metallicities. The temperature of the hot halo is comparable to the simplest virial temperature estimate of the galaxy, $T_{vir} = \frac{1}{2}\frac{\mu m_p}{k} v_c^2 = 3.6 \times 10^6$ K, (where $\mu \approx 0.6$ is the mean gas particle mass in units of the proton mass, $m_p$, and $v_c = 307$ km/s is the circular rotation velocity of the galaxy), and within the 1σ uncertainties in good agreement with the range of hot halo temperatures, $T \approx (4-6)\times 10^6$ K, predicted from detailed models (Toft et al. 2002) for a galaxy with circular velocity, $v_c \approx 300$ km/s.

The large extent of the X-ray halo of NGC 5746 in combination with the spectrum of the X-ray halo emission, the low star formation rate, and the non-detection of significant $H_\alpha$ emission outside the optical extent of the galaxy (quantified as $D_{25}$) in a 2 hour exposure with the Danish 1.54m telescope at La Silla (Rasmussen et al. 2005) is unprecedented for a quiescent spiral galaxy, demonstrating that here we are observing a new phenomenon. Below we compare the properties of the hot halo of NGC 5746 to expectations for out-flowing interstellar gas, heated by either supernovae or an active galactic nucleus, and to models with gradually in-flowing low-metallitic gas.

## 4. The origin of the NGC 5746 X-ray: Disc outflow vs. in-fall models

Spiral galaxies where large scale off-disc X-ray emission has been detected are typically starburst galaxies where the off-disc X-ray emission is always accompanied by optical line emission of similar spatial extent (Strickland et al. 2004a), believed to originate from gas ionised by UV radiation from massive and hence recently formed stars. For these spiral galaxies off-disc X-ray emission has been interpreted as originating from outflows powered by supernovae. However, the efficiency of supernova heating seen in other spiral galaxies is not sufficient to power the hot halo of NGC 5746. Spiral galaxies with detected off-disc X-ray emission have halo X-ray to far infrared luminosity ratios, $L_X / L_{FIR} = (2.2-8.0)\times 10^{-5}$, largely independent on the galaxy circular velocity (Strickland et al. 2004b) (where the halo X-ray luminosity is in the 0.3-2 keV band measured above 2 kpc from the disc), and with typical ratios for non-starburst galaxies, $L_X / L_{FIR} \lesssim 2.5\times10^{-5}$. The supernova rate is proportional to the far infrared luminosity (Heckman et al.1990) so if the off-disc X-ray emission is due to supernova heated gas, the X-ray to far infrared luminosity ratio is measuring the efficiency of converting supernova mechanical energy into X-rays. For NGC 5746 we find a halo ($|z| > 2$ kpc) 0.3-2 keV luminosity of $L_X = 7.6^{+5.1}_{-2.6}\times10^{39}$ erg/s, resulting in $L_X / L_{FIR} = 4.7\times10^{-4}$. Hence, in order to power its hot halo, supernova heating must be an order of magnitude more efficient in the quiescent galaxy NGC 5746 than in other spiral galaxies, and at least a factor twenty more efficient than in less massive, quiescent spiral galaxies, a very unlikely scenario. Furthermore, it is highly questionable whether supernovae are able to even blow out hot gas from the disc of NGC 5746 since its average supernova surface density rate over the disc is an order of magnitude lower than the corresponding disc supernova surface density rates in spiral galaxies with off-disc X-ray emission (Strickland et al. 2004b).

Irrespective of the energy source of a hypothetical outflow, the minimum energy input required to set up the NGC 5746 hot halo is $E_{W,\min} = M_h \Delta\varphi \approx 10^{57} - 10^{58}$ erg, where $M_h$ is the mass of the hot halo, and $\Delta\varphi$ is the galaxy potential difference between the pre-outflow site of the gas ($z_d < 5$ kpc) and the size of the hot halo ($z_h > 10$ kpc). Releasing this much energy in a starburst involves formation of $M_{*burst} \approx (5-10)\times10^9$ $M_\odot$ stars. In order to be consistent with the optical broad-band colours of NGC 5746, such a hypothetical starburst would have to be at least 1-2 Gyr old (using the GALAXEV code of Bruzual & Charlot 2003). However, the cause of such a massive starburst is rather mysterious since NGC 5746 is isolated and shows no signs of previous interactions with other galaxies. Furthermore, NGC 5746 shows no indications of an active galactic nucleus (AGN). A previous outburst from a presently "dormant" AGN would have to be as powerful as those seen in central cluster galaxies (Bîrzan et al. 2004) in order to create the NGC 5746 hot halo. However, radio observations (Condon 1987) (at 1.4

GHz to a limit of 0.1 mJy) have neither revealed a central source nor any off-disc emission (radio emitting "X-ray cavities") as seen in galaxy clusters with previous AGN outbursts of similar power. Furthermore, in our X-ray data we detect only a weakly obscured, faint source, $L_X \approx 2\times10^{40}$ erg/s, at the center of NGC 5746, orders of magnitude below the X-ray luminosity even for "dormant" central cluster AGN. Hence, the radio and X-ray data strongly indicate that any putative AGN in NGC 5746 is unlikely to have created its hot halo.

On the other hand, the X-ray properties of low-metallicity gas gradually flowing into the galaxy potential (at radial speeds of $\sim 10-20$ km/s at $r \sim 20$ kpc), deduced from recent numerical simulations of galaxy formation (Toft et al. 2002, see also Sommer-Larsen et al. 2005, Romeo et al. 2005), match those of the hot halo of NGC 5746. These simulations were set up mainly to reproduce the optical and cold gas characteristics of present day spiral galaxies, but they also predict the hot halo X-ray properties (with no additional adjustable parameters). The measured X-ray luminosity and surface brightness profile of the NGC 5746 hot halo are in excellent agreement with the predictions from the simulations; as is our non-detection of a hot halo around the less massive galaxy NGC 5170, see Fig. 3. Furthermore, the off-disc X-ray luminosities from the three (less massive) quiescent, edge-on spiral galaxies derived by Tüllmann et al. (2005) appear to be consistent with the predictions from numerical simulations (Fig. 3). For the two galaxies where Tüllmann et al. detect off-disc X-ray emission, their quoted X-ray luminosities are larger than halo luminosities given in Fig. 3 left (for the galaxies' respective circular velocities). However, Tüllman et al. extracted off-disc emission from regions much closer to the disc than in this study, increasing their off-disc X-ray luminosities relative to Fig. 3 left substantially (see Fig. 3 right). A detailed comparison between simulations and observations requires that exactly the same physical halo extraction region is used so this will have to await a future study.

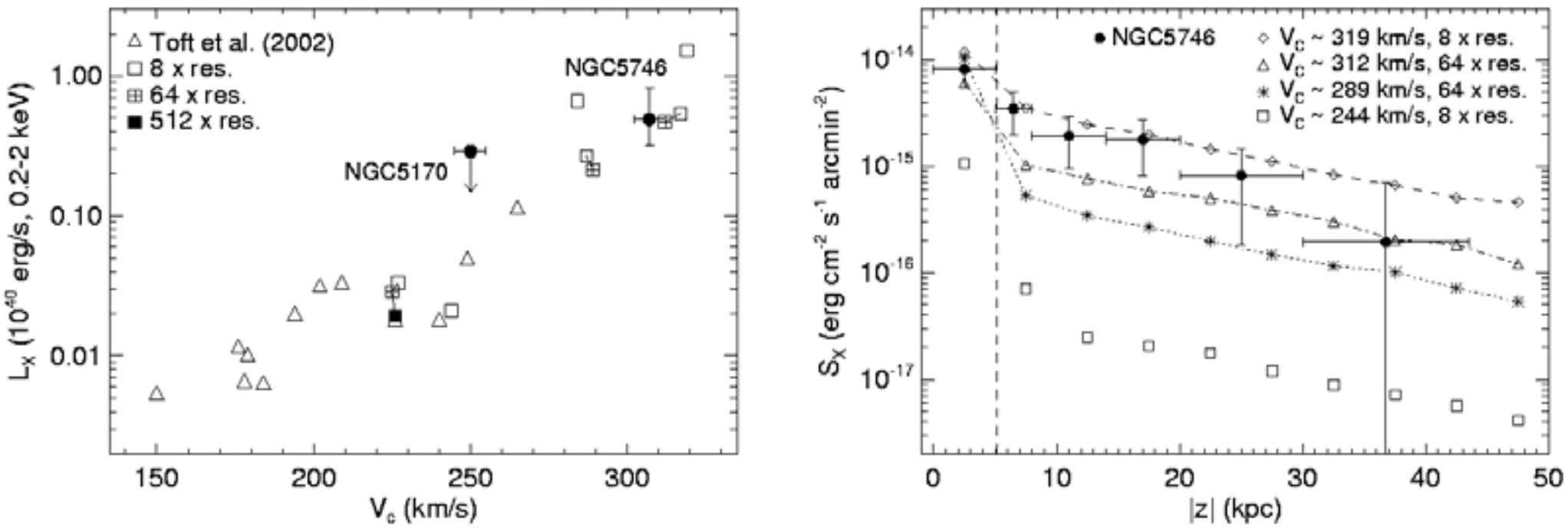

**Figure 3** Left: Predicted and observed 0.3-1.5 keV luminosities of X-ray haloes as a function of disc circular velocity. All X-ray luminosities have been calculated within the same physical aperture as used for NGC 5746. The filled circles are from the observations of NGC 5746 and NGC 5170 (1 sigma upper limit of $L_X < 2.9 \times 10^{39}$ erg/s , 0.3-2 keV) while other symbols are the predictions from simulations with a range of different resolutions and circular velocities: Triangles are for simulations (Toft et al. 2002) with primordial chemical composition while squares are for simulations with self-consistent chemical evolution (Sommer-Larsen et al. 2005, Romeo et al. 2005) run at 8, 64, and 512 times the original resolution (Toft et al. 2002) corresponding to a gas particle mass of $(75,\ 9.4, \text{and } 1.2) \times 10^4\ M_\odot$ respectively. Results from simulations re-run at higher resolution are connected with lines. Open squares, except the simulated galaxy with a circular velocity of ≈225 km/s, are for simulations run with a universal baryon fraction of 0.15. All other simulations were run with a baryon fraction of 0.1. Right: Predicted and observed surface brightness profile of X-ray haloes as function of the distance to the disc mid-plane. Filled circles are NGC 5746 data while other symbols mark simulations with different resolutions and circular velocities. The vertical dashed line indicates $D_{25}$ of NGC 5746.

Simple semi-analytical galaxy formation models (e.g. Benson et al. 2000, Toft et al. 2002) over-predict disc galaxy halo X-ray halo luminosities by more than an order of magnitude at z=0. This is due to the simplifying assumptions (shown by cosmological galaxy formation simulations to be incorrect, e.g. Toft et al. 2002, Keres et al. 2005) under-lying these models: The dark matter potential is assumed to be static, and the gas to comprise an infinite reservoir, which is in place and tracing the dark matter from the onset of galaxy formation. Subsequently, the gas is assumed to instantly cool and be deposited onto the disc from the "cooling radius", which may be hundreds of kpc at z=0. This is in marked contrast to simulations where the dark matter halo is very dynamic and building up gradually. Only gas within approximately the virial radius is available for cooling in a Hubble time, and most of the cooling is taking place fairly close to the disc (i.e. within $\sim 20-40$ kpc ); not at the cooling radius. All these effects reduce the predicted present day X-ray halo luminosity considerably. (Note that earlier epochs, disc galaxy halo X-ray luminosities were much larger, see Rasmussen et al., 2004). As a result, our predictions are in excellent agreement with the observed X-ray halo luminosities of NGC 5746 and NGC 5170, while simple analytical models (Benson et al. 2000) fail by a large margin.

## 5. Conclusions and outlook

Given the excellent match between observations and in-fall models, and the implausibility of alternative mechanisms discussed previously, the only compelling origin of the X-ray halo of NGC 5746 is thus that it is due to hot, probably shock heated, gas cooling radiatively as it descends into the galaxy's potential. The failure so far to detect such in-flowing hot gas around spiral galaxies hinges mainly on (a) the fact that the hot halo X-ray luminosity of NGC 5746, and presumably of spiral galaxies in general (Toft et al. 2002) is more than an order of magnitude lower than anticipated from predictions of simple semi-analytical galaxy formation models (Benson et al. 2000, Toft et al. 2002), and (b) that no massive edge-on spiral galaxies, like NGC 5746, has been targeted before (the predicted relation between halo X-ray luminosity and circular rotation velocity for massive galaxies is very steep, $L_X \propto v_c^7$ , see Fig.3).

The present detection of the long-sought X-ray halo around quiescent spiral galaxies like NGC 5746 strongly indicates that (at least) massive galaxies are able to retain a hot gaseous halo to the present day. Some of this halo gas may eventually cool out and be deposited onto the disc, acting as a supply of fresh material for continuous star formation. Hence, we are here likely witnessing the on-going galaxy formation process, in line with hierarchical galaxy formation models (White & Rees 1978). One of the predictions in this scenario, that the metallicity of the hot halo gas is low ($[Fe/H] \lesssim -1$), should be directly testable through very deep X-ray spectroscopy.

## Acknowledgements

KP and JR acknowledge support from the Instrument Center for Danish Astrophysics and the Danish Natural Sciences Research Council. KP acknowledges support from the Carlsberg Foundation. JSL acknowledges support from the Villum Kann Rasmussen Foundation. ST acknowledges support from the Danish Natural Sciences Research Council. AJB acknowledges support from a Royal Society University Research Fellowship. RGB acknowledges the support of a PPARC Senior Fellowship. The Dark Cosmology Centre is funded by the DNRF. This work was based on observations with the Chandra X-ray Observatory, and with the Danish 1.54m telescope at the La Silla Observatory. The simulations reported here were all run on DCSC supercomputers; abundant access to these facilities is greatfully acknowledged.